\documentclass[12pt]{article}%
\usepackage{amsmath}
\usepackage{amsfonts}
\usepackage{amssymb}
\usepackage{graphicx}%
\providecommand{\U}[1]{\protect\rule{.1in}{.1in}}
\newcommand{\bfr}{\begin{flushright}}
\newcommand{\efr}{\end{flushright}}
\newcommand{\bc}{\begin{center}}
\newcommand{\ec}{\end{center}}
\newcommand{\ben}{\begin{enumerate}}
\newcommand{\een}{\end{enumerate}}

\newcommand{\be}{\begin{equation}}
\newcommand{\ee}{\end{equation}}
\newcommand{\ba}{\begin{array}}
\newcommand{\ea}{\end{array}}

\def\6{\partial}
\newcommand{\bea}{\begin{eqnarray}}
\newcommand{\eea}{\end{eqnarray}}
\begin{document}

\title{Semiclassical bosonic D-brane boundary states in curved spacetime}
\author{Ion V. Vancea\thanks{email:ionvancea@ufrrj.br}}
\date{25 February 2008}
\maketitle

\pagestyle{empty}

\begin{center}
\emph{Departamento de F\'{\i}sica, Universidade Federal Rural do Rio de
Janeiro (UFRRJ),\newline Cx. Postal 23851, 23890-000 Serop\'{e}dica - RJ,
Brasil}
\end{center}

\abstract{We give a simple method to calculate the semiclassical $D$-brane boundary states of the bosonic string propagating in an arbitrary curved spacetime in a perturbative approach in which the metric $g_{AB}$ and the constant antisymmetric Kalb-Ramond field $b_{AB}$ form the general background which is treated exactly. As an important particular case, it is shown that at the first order perturbation theory there are $D$-brane coherent states in the $d$-dimensional {\em AdS} spacetime if certain conditions are fulfilled by the boundary projectors.}

\newpage

\section{Introduction}

The $D$-brane dynamics has been proved useful to understanding a large class of physical phenomena such as the black hole structure, the string cosmologies, the early Universe, the duality between the Quantum Field Theory  and the General Relativity, the primordial gravitational waves - to mention just few of the most studied problems nowdays. In all these cases, and in many others, the string theory undelying the $D$-branes interacts with a strong gravitational field and background massless fields in a general spacetime manifold. In the flat spacetime and in the constant Kalb-Ramond (KR) field, it is possible to give at least two different interpretations to the $D$-brane: the first one is the (classical) geometric hyperplane on the boundary of the string worldsheet and the second one is the (quantum) coherent state constructed out of states that belong to the closed string Fock space \cite{jpbook}. There is a strong relation between these two pictures as pointed out in \cite{vfp}. On the other hand, there is no quantum description
of the $D$-brane in general curved manifolds. Therefore, most of the calculations involving $D$-branes in non-trivial backgrounds rely on solutions of the low energy limit string theory; quantum effects may not be apparent in this
treatement. Since a consistent quantum field theory in arbitrary spacetime is not available yet, it is not possible to construct an exact quantum $D$-brane model in the most general case. The aim of this letter is to provide a simple method to construct perturbative bosonic {\em D}-brane boundary states in curved spacetime and in the presence of a constant KR-field in which the background fields are maintained exact. As an important application, we determine the conditions for the existence of the {\em D}-brane coherent states at the first order perturbation (strong gravitational field) in the $d$-dimensional {\em AdS} spacetime.

As is well known, expanding the metric of a strong gravitational field arround the flat spacetime leads to an inaccurate theory. Therefore, we are going to apply a semiclassical quantization scheme developed to analyse the bosonic
string quantum dynamics in curved spacetime in \cite{ns1,ns2,ns3}. The main feature of this method is that the metric is treated exactly while the string fields are expanded arround the trajectory of the center of mass in the free
falling reference frame. By chossing a local light-cone gauge, the string degrees of freedom can be locally split into transversal to the geodesic, and longitudinal which are functions of the transversal ones. The transversal
degrees of freedom can be quantized order by order in the perturbation theory by using the canonical quantization methods. The dimensionless perturbation parameter is $\epsilon= \sqrt{\pi} l_{Planck}/R_{c}$ where $R_{c}$ is the
typical curvature radius of the background. The first order approximation is predominant in the case of strong gravitational field \footnote{For a review of the application of the semiclassical quantization method see \cite{ns4}. For more recent applications to coherent states and string thermodynamics see \cite{ns6,ns7,ns8, ivv1,eg1}.}.

The paper is organized as follows. In Section 2 we give the semiclassical $D$-brane boundary conditions in a general curved spacetime in the background composed by an arbitrary metric and a constant KR-field. In Section 3 we obtain the conditions for the existence of the $D$-brane coherent states in strong gravitational field in the {\em AdS} spacetime. The last section is devoted to conclussions and discussions.

\section{Semiclassical {\em D}-brane boundary conditions}

Consider the bosonic string in a curved background in the presence of a KR-field. Let $(N,\gamma_{\alpha\beta})$, $\alpha,\beta=0,1$ be the two dimensional world-sheet of boundary $\partial N$ and $(M,g_{AB})$, $A,B=0,1,\ldots,d-1$, be the $d$-dimensional target space. The string coordinates are the components of the analytic embedding $\phi:N\hookrightarrow M$. The classical string dynamics is defined by the non-linear $\sigma$-model \cite{wz} 
\begin{equation}
S\left[  \gamma,\phi,g(\phi),b(\phi)\right]  =\frac{1}{2}%
{\displaystyle\int\nolimits_{N}}
d^{2}\sigma\sqrt{\gamma}\left[  \gamma^{\alpha\beta}(\sigma)g_{AB}(\phi
)+\frac{q}{2}\varepsilon^{\alpha\beta}b_{AB}(\phi)\right]  \partial_{\alpha
}\phi^{A}\partial_{\beta}\phi^{B}. 
\label{sigma-model}
\end{equation}
In order to derive the conditions for the existence of semiclassical $D$-brane boundary states, we take $N=S^{1}\times \mathbb{R}^{+}$ and pick up the conformal gauge $\gamma_{\alpha\beta}=\eta_{\alpha\beta}$. Also, we take the KR-field constant. Then the equations of motion, the energy-momentum tensor and the Virasoro conditions are given by the following relations
\begin{align}
\ddot{\phi}^{A} &  -\phi^{A^{\prime\prime}}+\Gamma_{BC}^{A}\left(
\phi\right)  \left(  \dot{\phi}^{B}\dot{\phi}^{C}-\phi^{B^{\prime}}%
\phi^{C^{\prime}}\right)  =0,
\label{eqs-motion}\\
T_{\alpha\beta} &  =g_{AB}\partial_{\alpha}\phi^{A}\partial_{\beta}\phi
^{B}-\frac{1}{2}\eta_{\alpha\beta}\left(  \eta^{\lambda\delta}g_{AB}%
+q\varepsilon^{\lambda\delta}b_{AB}\right)  \partial_{\lambda}\phi^{A}%
\partial_{\delta}\phi^{B},
\label{en-mom-tensor}\\
g_{AB}\dot{\phi}^{A}\phi^{B^{\prime}} &  =g_{AB}\left(  \dot{\phi}^{A}%
\dot{\phi}^{B}+\phi^{A^{\prime}}\phi^{B^{\prime}}\right)  =b_{AB}\dot{\phi
}^{A}\phi^{B^{\prime}}=0,
\label{virasoro-constr}%
\end{align}
where $\dot{\phi}^{A}=\partial_{\tau}\phi^{A}$, $\phi^{B^{\prime}} =\partial_{\sigma}\phi^{B}$, $\tau\in%
\mathbb{R}^{+}$ and $\sigma\in\left[  0,\pi\right]$. The relations (\ref{eqs-motion}), (\ref{en-mom-tensor}) and (\ref{virasoro-constr}) hold only if the boundary terms on $\6 N$ vanish. This is achieved by imposing either Dirichlet or Neumann boundary conditions on each string field $\phi^A$. Different $D$-branes are defined by mixtures of Dirichlet and Neumann boundary conditions which can be written in terms of the projectors 
$\left( P_{\pm}\right)_{AB}= \frac{1}{2}\left(g_{AB}\pm R_{AB}\right)$ \cite{alz1,alz2,kns}
\be
\left(  P_{+} \right)_{B}^{A}\left(  \phi^{B^{\prime}}-q b_{C}^{B}\dot{\phi}^{C} \right)|_{\partial N} =0,\qquad
\left(  P_{-} \right)_{B}^{A}\delta \phi^{B}|_{\partial N} =0, 
\label{D-brane-bc}
\ee
where the matrices $R_{AB}$ satisfy the following relations  
\begin{equation}
R_{C}^{A}R_{B}^{C}=\delta_{B}^{A},\qquad g_{AB}R_{C}^{A}R_{D}^{B}=g_{CD}.
\label{R-property}
\end{equation}
In curved background, the free string moves along the trajectory of its center of mass. That is a geodesic $\phi_{0}^{A}(\tau)$ that fulfills the following set of equations
\be
\ddot{\phi}_{0}^{A}(\tau) + \Gamma_{BC}^{A}(\phi_{0})\dot{\phi}_{0}^{B}(\tau)\dot{\phi}_{0}^{C}(\tau)=0,\qquad
g_{AB}(\phi_{0})\dot{\phi}_{0}^{A}\dot{\phi}_{0}^{B}=-m^{2}\alpha^{\prime2}, 
\label{cm-relations}
\ee
where the parameter of the geodesic is chosen to be the variable $\tau$. Following \cite{ns1,ns2,ns3}, we perform the
expansion of the string fields $\phi^{A}(\tau , \sigma )$ in powers of $\epsilon$ 
\begin{equation}
\phi^{A}\left( \tau,\sigma \right) = \sum_{n=0}^{\infty}\epsilon^{n}\phi_{n}^{A}\left( \tau,\sigma\right), 
\label{power-ex}
\end{equation}
with the boundary condition $\phi_{0}^{A} \left( \tau,\sigma\right) = \phi_{0}^{A}\left( \tau \right)$. By interpreting the string coordinates as quantum fields and using the relations (\ref{D-brane-bc}) and (\ref{power-ex}), the perturbative {\em D}-brane boundary state $\left\vert B \right\rangle$ is defined at each order in $\epsilon$ as the state from the Hilbert space of the bosonic string that satisfies the following equations
\begin{align}
\left(  P_{+}\right)  _{B}^{A}b_{C}^{B}\dot{\phi}_{0}^{C}\vert_{\partial N}
\left\vert B\right\rangle  &  =0, \qquad
\left(  P\_\right)  _{B}^{A}\delta\phi_{0}^{B} \vert _{\partial N} \left\vert B\right\rangle  
=0, 
\label{zero-order}\\
\left(  P_{+}\right)  _{B}^{A}\left(  \phi_{i}^{B}-qb_{C}^{B}\dot{\phi}_{i}^{C}\right) \vert_{\partial N} 
\left\vert B\right\rangle  &  =0, \qquad 
\left(  P\_\right)  _{B}^{A}\delta\phi_{i}^{B}\vert_{\partial N} \left\vert B\right\rangle =0, \qquad i=1,2,\ldots
\label{i-th-order}
\end{align}
Some remarks are in order here. In general, the state $\left\vert B \right\rangle$ does not have a simple interpretation in terms of quantum string excitations since the string modes, beside their interaction with the background fields, interact among themselves. Nevertheless, the terms of the power expansion (\ref{power-ex}) 
are organized such that at the first order perturbation in $\epsilon$ there are only terms from the interaction of the string excitations and the background fields. The string Fock space is completely determined at the first order perturbation theory by the Virasoro conditions, and it contains only the states of the free transversal string excitations \footnote{Since in any non-inertial reference frame one can choose a local coordinate system, the longitudinal excitations can be expressed as functions of the transversal ones \cite{ns1}.}. Note that the first order approximation is predominant for large scalar curvature compared to the string scale. That is the case of, for example, the black-hole background with black-hole mass larger than string energy, and the cosmological backgrounds \cite{ns1,ns2}. 

Let us consider the strong gravitational field approximation. Since in the gravitational field the string is massive, one can introduce $d-1$ polarization vectors $n_{a}^{A}$ which satisfy the following relations 
\be
g_{AB}(\phi_{0})n_{a}^{A}\dot{\phi}_{0}^{B}  = 0,~~
g_{AB}(\phi_{0})n_{a}^{A}n_{b}^{B}    = \delta_{ab},~~ 
g^{AB} (\phi^{0}) = -\frac{1}{m^2\alpha^{\prime}} \dot{\phi}_{0}^{A}\dot{\phi}_{0}^{B}+n^{Aa}n^{B}_{a},
\label{normal-vect}
\ee
where $a,b = 1,2,\ldots,d-1$. The $SO(d-1)$ gauge symmetry of the polarization vectors can be fixed by choosing the covariant gauge $\dot{\phi}_{0}^{A}\nabla_{A}n_{b}^{B}= 0$. The first order string fields $\phi^{A}_{1}$ can be written in terms of polarization vectors
\be
\phi_{1}^{A}(\tau,\sigma)=\varphi^{a}(\tau,\sigma)n_{a}^{A},
\label{first-order-decomp}
\ee
where the normal string excitations $\varphi^{a}(\tau,\sigma)$ have the standard Fourier expansion
\be
\varphi^{a}(\tau,\sigma)= \sum_{k \in \mathbb{Z}}C_{k}^{a}(\tau)e^{-ik\sigma}.
\label{Fourier-ex-var}
\ee
The string operators $C_{ka}$ satisfy the following equation of motion
\be
\ddot{C}_{ka} (\tau )+\left(  k^{2}\delta_{ab}-R_{ABCD}n_{a}^{A} n_{b}^{B}
\dot{\phi}_{0}^{C}\dot{\phi}_{0}^{D}\right)  C_{k}^{b}(\tau)=0,
\label{eq-mot-C}
\ee
where $R_{ABCD}$ is the curvature tensor associated to the metric $g_{AB}$. By plugging
the relation (\ref{first-order-decomp}) into the equation (\ref{i-th-order}) one obtains the following set of equations
\begin{equation}
\left(  P_{+}\right)  _{B}^{A}\left(  k\delta_{C}^{B}+qb_{C}^{B}\partial_{\tau}\right)
n_{a}^{C}C_{k}^{a}(0)\left\vert B\right\rangle =0,~~
\left(  P\_\right)_{B}^{A}n_{a}^{B}\delta C_{k}^{a}(0)\left\vert B\right\rangle =0,~~
k\in \mathbb{Z}.
\label{first-order-bc}
\end{equation}
The equations (\ref{first-order-bc}) define the first order perturbation bosonic $D$-brane boundary states in a strong gravitational field and in the presence of a constant KR field on an arbitrary manifold $M$ in the
center of mass reference frame, written in terms of the string mode operators. Since we have perturbed the string fields around a geodesic, the above equations have a local character. In many backgrounds, the operators $C_{k}^{a}(\tau)$ have a simple interpretation in terms of creation and annihilation operators acting on the Fock space of the free string modes in interaction with the background metric. That occurs, for example, in the black-hole anti-de Sitter, de Sitter and anti-de Sitter backgrounds, to mention just few of them \cite{ns4}. In these cases we can look after $D$-brane states similar to those that exist in the flat spacetime. In what follows we are going to specialize our discussion to the important case of the \emph{AdS} spacetime.

\section{First order {\em D}-branes in {\em AdS}-spacetime}

The \emph{AdS} background does not support a conformal quantum string theory in arbitrary $d$ dimensions. However, the black-hole \emph{AdS} spacetime is an exact solution of the string theory in $d=2+1$ dimensions \cite{bhtz}. In the semiclassical approach, the conformal generators and the first order string fields are independent of the black-hole mass at the first order perturbation theory. That allows one to generalize the semiclassical quantization in the first order perturbation to the \emph{AdS} spacetime of arbitrary dimension \cite{ns5}. In this case, the Fourier expansion (\ref{Fourier-ex-var}) takes the following form 
\begin{align}
\varphi^{a}(\tau,\sigma) 
&  
=\sum_{k\neq0}\left(  \frac{2\left\vert k\right\vert \Omega_{k}}{\alpha^{\prime}}\right)^{\frac{1}{2}}
\left[ \alpha_{k}^{a}e^{-ik\left(  \Omega_{k}\tau-\sigma\right)  }+
\beta_{k}^{a}e^{-ik\left(  \Omega_{k}\tau+\sigma\right)  }\right] 
\nonumber\\
& + \left(  \frac{l}{2m}\right)  ^{\frac{1}{2}}\left[  \alpha_{0}^{a} e^{-\frac{im\alpha^{\prime}}{l}\tau}+\beta_{0}^{a}e^{\frac{im\alpha^{\prime}}{l}\tau}\right],
\label{first-ord-fourier}
\end{align}
where, $l=H^{-1}$ and $H$ is related with the cosmological constant by
$\Lambda=-(d-1)(d-2)H^{2}/2$.  The perturbation parameter in the \emph{AdS} spacetime is $\epsilon=\alpha^{\prime2}l$. The factors $\Omega_{k}=\Omega_{-k}$ are defined as
\begin{equation}
\Omega_{k}=\sqrt{ 1+\frac{m^{2}\alpha^{\prime2}}{k^{2}l^{2}}}.
\label{omega-factor}
\end{equation}
The relation (\ref{first-ord-fourier}) represents the decomposition of the
first order string excitations in terms of transverse oscillators of
frequencies $\omega_{k}=\left\vert k\right\vert \Omega_{k}$, for $k\neq0$ and
$\omega_{0}=m\alpha^{\prime}/l$. The operators $\alpha_{k}^{a}$ and $\beta
_{k}^{a}$ are creation and annihilation operators of the right- and left-modes
for $k<0$ and $k>0$, respectively. Indeed, one can easily show that they satisfy the
following relations
\begin{align}
\alpha_{k}^{a} &  =\left(  \alpha_{-k}^{a}\right)  ^{\dag},\qquad\beta_{k}%
^{a}=\left(  \beta_{-k}^{a}\right)  ^{\dag},\qquad\beta_{0}^{a}=\left(
\alpha_{0}^{a}\right)  ^{\dag},
\label{comm-1}\\
\left[  \alpha_{k}^{a},\alpha_{s}^{b\dag}\right]   &  =\left[  \beta_{k}%
^{a},\beta_{s}^{b\dag}\right]  =\delta^{ab}\delta_{ks},
\label{comm-2}\\
\left[  \alpha_{k}^{a},\beta_{s}^{b}\right]   &  =0,\qquad\left[  \alpha
_{0}^{a},\alpha_{0}^{b\dag}\right]  =\delta^{ab},
\label{comm-3}
\end{align}
for all $k,s>0$. As in the flat spacetime, one can define the physical Hilbert space of the
first order free excitations by imposing the quantum Virasoro conditions
$L_{k}^{\alpha}\left\vert \psi_{phys}\right\rangle =L_{k}^{\alpha}\left\vert
\psi_{phys}\right\rangle =0$, $\forall k\in
\mathbb{Z}_{+}^{\ast}$ \ and $\left(  L_{0}^{\alpha}-2\pi\alpha^{\prime}a\right)
\left\vert \psi_{phys}\right\rangle =\left(  L_{0}^{\beta}-2\pi\alpha^{\prime
}a\right)  \left\vert \psi_{phys}\right\rangle =0$ for zero modes. These
relations should be supplemented by the level matching condition which takes
the following form
\begin{equation}
\pi\alpha^{\prime}\sum_{k>0}k\sum_{a=1}^{d-1}\left(  \beta_{k}^{a\dag}
\beta_{k}^{a}-\alpha_{k}^{a\dag}\alpha_{k}^{a}\right)  \left\vert \psi
_{phys}\right\rangle =0.
\label{level-matching}
\end{equation}
The operators $L_{k}^{\alpha}$ and $L_{k}^{\beta}$ are the Virasoro operators.
They are defined as the Fourier coefficients of the energy-momentum components
$T_{\pm\pm}$ in the worldsheet light-cone coordinates as in the flat
spacetime. The operators $L_{k}^{\alpha}$ and $L_{k}^{\beta}$ act on the Fock
space constructed from the vacuum state $\left\vert 0\right\rangle $ that is
annihilated to zero by all creation operators \cite{ns5}. 

In order to obtain the first order $D$-brane boundary conditions, one has to plug the relation (\ref{first-ord-fourier}) into the equation (\ref{first-order-bc}). After some calculations, one arrives at the following boundary state equations in the Fock space of the string perturbations
\begin{align}
\left(  u_{ka}^{A}\alpha_{k}^{a}+v_{ka}^{A}\beta_{k}^{a\dag}\right)  \left\vert
B\right\rangle  &  =0,
\label{AdS-bc-1}\\
\left(  u_{ka}^{A}\alpha_{k}^{a\dag}+v_{ka}^{A}\beta_{k}^{a}\right)  \left\vert
B\right\rangle  &  =0,
\label{AdS-bc-2}\\
w_{a}^{A}\left(  \beta_{0}^{a}-\alpha_{0}^{a}\right)  \left\vert
B\right\rangle  &  =0. 
\label{AdS-bc-3}
\end{align}
where $k>0$. Here, we have used the following notations
\begin{align}
\lambda_{kB}^{A}  &  =\delta_{B}^{A}+q\Omega_{k}b_{B}^{A},\quad\widetilde
{{\lambda}}_{kB}^{A}=\delta_{B}^{A}-q\Omega_{k}b_{B}^{A},
\label{not-1}\\
u_{ka}^{A}  &  =\left(  P_{+}\right)  _{B}^{A}\lambda_{kC}^{B}n_{a}^{C},\quad
v_{ka}^{A}=\left(  P_{+}\right)  _{B}^{A}\widetilde{\lambda}_{kC}^{B}n_{a}
^{C},\quad w_{a}^{A}=\left(  P_{+}\right)  _{B}^{A}b_{C}^{B}n_{a}^{C}.
\label{not-2}
\end{align}
The equations (\ref{AdS-bc-1}), (\ref{AdS-bc-2}) and (\ref{AdS-bc-3}) represent the boundary state equations in the Fock space of the first order string perturbations in strong gravitational field and in the presence of a constant KR-field (torsion field) in the {\em AdS} spacetime. The boundary state equations are expressed in terms of free string oscillation modes. Therefore, it is interesting to see whether they admit $D$-brane coherent state solutions. 

Consider the following ansatze
\begin{equation}
\left\vert B\right\rangle =\exp\left(  g\sum_{a=1}^{d-1}\sum_{b=1}^{d-1}%
\delta_{ab}\alpha_{0}^{a\dag}\alpha_{0}^{b\dag}\right)  \exp\left(
f\sum_{k \neq 0}\sum_{a,b=1}^{d-1}\sum_{A,B=0}^{d-1}\alpha_{k}^{a\dag}u_{ka}%
^{A}g_{AB}v_{kb}^{B}\beta_{k}^{b\dag}\right)  \left\vert 0\right\rangle
,\label{ansatze-coh-state}%
\end{equation}
where $g,f$ are constant phase factors and $\left\vert B\right\rangle $ is defined up to a normalization constant. A little algebra shows that the above ansatze is a solutions of the boundary equations (\ref{AdS-bc-1}) and (\ref{AdS-bc-2}) only if $f=-1$ and
\begin{equation}
u_{ka}^{A}u^{aB}_{k}=v_{ka}^{A}v^{aB}_{k}=g^{AB}.
\label{constraints}%
\end{equation}
By using the relations (\ref{normal-vect}), (\ref{not-1}) and (\ref{not-2}),
one can express the constraints (\ref{constraints}) as relations among the
components of the projector $(P_{+})_{AB}$ that defines the $D$-brane. 
The final relation has the following form
\begin{align}
& \left(  P_{+}\right)  _{B}^{A}\left[  \left(  P_{+}\right)  _{D}^{C}\left(
\dot{\phi}_{0}^{B}\dot{\phi}_{0}^{D}+q^{2}\Omega_{k}^{2}b_{E}^{B}b_{F}%
^{D}\left(  \dot{\phi}_{0}^{E}\dot{\phi}_{0}^{F}+m^{2}\alpha^{\prime2}%
g^{EF}\right)  \right)  +m^{2}\alpha^{\prime2}\left(  P_{+}\right)  _{D}%
^{B}g^{DC}\right]  \nonumber\\
& =m^{2}\alpha^{\prime2}g^{AC},\quad\forall k>0.
\label{constraints-P}%
\end{align}
The zero mode part of $\left\vert B\right\rangle$ can be calculated by using the equation (\ref{AdS-bc-3}). 
It follows that $g=1/2$. 

In conclussion, there are $D$-brane coherent states of the form (\ref{ansatze-coh-state}) in the $AdS$ spacetime at the first order in the perturbation theory provided that the corresponding projectors satisfy the equation (\ref{constraints-P}).

\section{Conclussions and Discussions}

We have used the semiclassical quantization method to derive the perturbative
$D$-brane boundary conditions in curved spacetime and in the presence of a
constant KR-field. The strong gravitational field corresponds to the first
order perturbation theory for which the boundary equations in arbitrary
background metric have been derived in the equations (\ref{zero-order}) and
(\ref{i-th-order}). In the \emph{AdS} background, the first order boundary
conditions have been written in the Fock space of the free transversal string
modes in (\ref{AdS-bc-1})-(\ref{AdS-bc-3}). We have shown that there are
$D$-brane coherent states that satisfy these equations if the projectors that
define the $D$-brane fulfill the equation (\ref{constraints-P}).

Let us discuss some particular cases of the equations (\ref{constraints-P}).
In the torsionless \emph{AdS} background, the zeroth order string coordinates
$\dot{\phi}_{0}^{A}$'s have simple relationships with the matrix $R_{AB}$.
Indeed, if there is a tensor $\Pi_{CD}^{AB}$ that satisfies the relation
\begin{equation}
\Pi_{CD}^{AB}\left(  \delta_{A}^{E}\delta_{B}^{F}+2R_{A}^{E}\delta_{B}%
^{F}+R_{A}^{E}R_{B}^{F}\right)  =\delta_{C}^{E}\delta_{D}^{F}%
,\label{torsionless}%
\end{equation}
then it follows from (\ref{constraints-P}) that
\begin{equation}
\dot{\phi}_{0}^{A}\dot{\phi}_{0}^{B}=2m^{2}\alpha^{\prime2}\left[  \Pi\left(
g-R\right)  \right]  ^{AB}. \label{torsionless-R}%
\end{equation}
In general, $\dot{\phi}_{0}^{A} \neq 0$ from which it follows that $\alpha^{\prime} \neq 0$.
On the other hand, in the particle limit $T\rightarrow\infty$ the equation
(\ref{constraints-P}) is reduced to
\begin{equation}
R_{C}^{B}V_{k}^{CD}=V_{k}^{BD}-\dot{\phi}_{0}^{B}\dot{\phi}_{0}^{D}%
-q^{2}\Omega_{k}^{2}b_{E}^{B}b_{F}^{D}\dot{\phi}_{0}^{E}\dot{\phi}_{0}%
^{F},\quad\forall k>0,\label{particle-limit}%
\end{equation}
where $V_{k}^{AB}$ are the components of arbitrary tensors $V_{k}$ that should
depend on $k$. The integrability of the projector $P_{+}$ expressed by the
relation \cite{kns}
\begin{equation}
\left(  P_{+}\right)  _{[B}^{A}\left(  P_{+}\right)  _{D]}^{C}\left(
P_{+}\right)  _{A,C}^{E}=0,
\label{integrability-P}%
\end{equation}
imposes the following constraints on the tensors $V_{k}$%
\begin{equation}
V_{kD}^{[A}\left(  P_{+}\right)  _{B}^{C]}\left(  P_{+}\right)  _{A,C}%
^{E}=\frac{1}{2}\left[  \dot{\phi}_{0D}\dot{\phi}_{0}^{[A}\left(
P_{+}\right)  _{B}^{C]}+q^{2}\Omega_{k}^{2}\dot{\phi}_{0}^{M}\dot{\phi}%
_{0}^{N}b_{ND}b_{M}^{[A}\left(  P_{+}\right)  _{B}^{C]}\right]  \left(
P_{+}\right)  _{A,C}^{E},
\label{constr-V-tensor}%
\end{equation}
for all $k>0$. Note that the limit of vanishing KR-field can be safely
taken in the equations (\ref{particle-limit}) and (\ref{constr-V-tensor}).

\noindent\textbf{Acknowledgments} I would like to thank to J. A. Helay\"{e}l-Neto for discussions and to S. A. Dias and A. M. O. de Almeida for hospitality at LAFEX-CBPF where part of this work was accomplished.

\end{document}